# Event Analysis of Pulse-reclosers in Distribution Systems Through Sparse Representation

M. Ehsan Raoufat, *Student Member, IEEE,* Ali Taalimi, *Student Member, IEEE,* Kevin Tomsovic, *Fellow, IEEE,* and Robert Hay

*Abstract*—The pulse-recloser uses pulse testing technology to verify that the line is clear of faults before initiating a reclose operation, which significantly reduces stress on the system components (e.g. substation transformers) and voltage sags on adjacent feeders. Online event analysis of pulse-reclosers are essential to increases the overall utility of the devices, especially when there are numerous devices installed throughout the distribution system. In this paper, field data recorded from several devices were analyzed to identify specific activity and fault locations. An algorithm is developed to screen the data to identify the status of each pole and to tag time windows with a possible pulse event. In the next step, selected time windows are further analyzed and classified using a sparse representation technique by solving an $\ell_1$-regularized least-square problem. This classification is obtained by comparing the pulse signature with the reference dictionary to find a set that most closely matches the pulse features. This work also sheds additional light on the possibility of fault classification based on the pulse signature. Field data collected from a distribution system are used to verify the effectiveness and reliability of the proposed method.

## I. Introduction

The electrical distribution system is critical for reliable delivery of power but is prone to disturbances or electrical faults [1]. Power interruptions due to faults in distribution network cost millions of dollars annually for utilities and their customers. Around 75% - 90% of the total number of faults are temporary lasting for a few cycles and can be cleared by a recloser [2]. Usually these temporary faults occur when phase conductors momentarily contact other phases or connect to the ground due. However, in case of a permanent fault, a conventional recloser stresses the equipment with fault current and produces voltage sag every time it switches after an event. One solution is to use pulse-reclosers to inject a low-energy test pulse into the line before attempting a reclose. The term pulse-recloser used in this paper as a general term for switching devices with pulse testing capabilities.

IntelliRupter PulseCloser is an example of a pulse-recloser device with PulseClosing technology[1], which significantly decreases stress on the system components (e.g. substation transformers, lines, etc.), eliminates conductor slaps and reduces voltage sag on adjacent feeders [3]. This device is a unique alternative to conventional automatic reclosers and has the ability to work in stand-alone mode as a fault interrupter. These devices can also be used for fault isolation in distribution system and islanding operation in smart grids [4], [5].

Both sides of the device are equipped with accurate voltage and current sensors that provide high resolution (64 samples/cycle) time-stamped event data [6]. Recorded data can then be used for fault classification, estimation of the fault location and to verify proper operation. In particular, the current and voltage pulse waveforms contain important information. For real-time applications, time domain analysis can be used to provide basic information and screen the data to find windows of possible pulses. It is sometimes difficult to detect these pulses as different type of loads, faults or transformers may influence the pulse amplitude or duration. In this paper, we design a dictionary learning framework based on matrix factorization algorithm to represent time domain pulses in the space of sparse codes which leads to a higher classification accuracy.

There have been previous works on event analysis of capacitor banks [7] and protection relays [8] in distribution systems. However, to the best of our knowledge, this is the first work on event analysis of pulse-reclosers. Our contribution is two-fold. First, we consider time domain analysis to identify the status of each pole and tag time windows with a possible pulse event. Second, we extract a discriminative representation of pulses which leads to a high classification accuracy. The idea is to learn a dictionary that transfers data from the original time domain to the space of sparse codes such that is favorable towards the maximal discriminatory power.

A dictionary in signal processing can be simplify as a set of fixed variables and then seeks for the solution as a linear combination of variables in the dictionary. The dictionary should be designed so that it can successfully generalize unseen and new data [9]. The prior information about the data or the form of the solution leads to the concept of regularization, which shows promising performance in dealing with unseen data. In this paper, sparse solutions are preferred, which leads to $\ell_1$-norm regularization. This bias toward sparsity lies in seeking a simple reasoning for the task that is easy to interpret and has a low processing complexity.

The rest of this paper is organized as follows: Section II discusses the operation of pulse-reclosers and basic detection rules. In Section III, pulse signature dictionaries have been

This work was supported in part by the National Science Foundation under grant No CNS-1239366, and in part by the Engineering Research Center Program of the National Science Foundation and the Department of Energy under NSF Award Number EEC-1041877 and the CURENT Industry Partnership Program.

M. Ehsan Raoufat, Ali Taalimi and Kevin Tomsovic are with the Min H. Kao Department of Electrical Engineering and Computer Science, The University of Tennessee, Knoxville, TN 37996 USA (e-mail: mraoufat@utk.edu).

Robert Hay is with Electric Power Board Chattanooga, TN 37404.

[1]IntelliRupter, PulseCloser and PulseClosing are registered trademarks of S&C Electric Company.

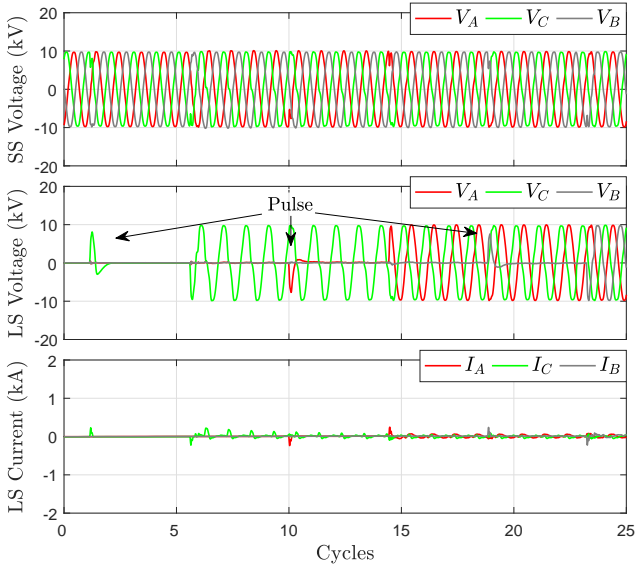

Fig. 1. Response to a temporary fault.

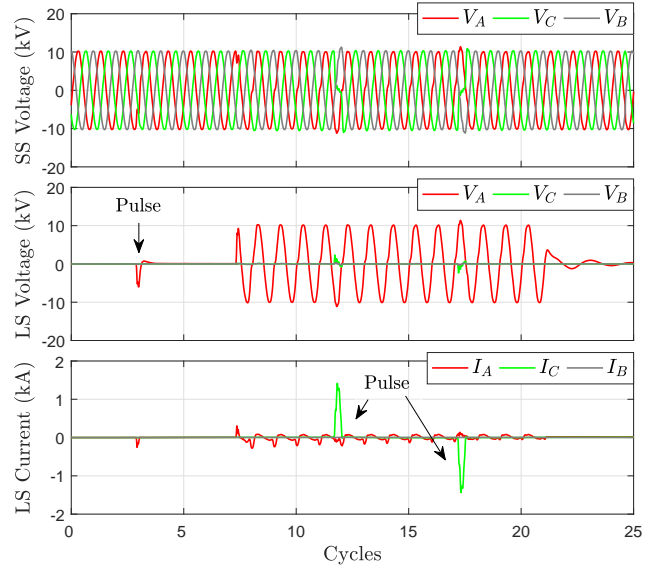

Fig. 2. Response to a permanent fault in phase C.

constructed and then sparse representation technique for pulse detection has been explained. Section IV conducts the performance evaluation using actual field data and results have been discussed. Finally, conclusions are presented in Section V.

## II. OPERATION AND DATA ANALYSIS OF PULSE-RECLOSERS

Pulse testing technology has become available as an integrated package for overhead and underground applications [6]. These devices are becoming increasingly popular in distribution system and transferring the huge volume of data into useful information is an important task to fully take advantage of the devices. The following reviews the basic operation of pulse-reclosers.

### A. Pulse Testing Operation

Pulse testing operation initiates a three-phase pulse and inverse-pulse test of each pole separately to evaluate the line status. This is accomplished by a sub-cycle close-open operation of switchgear contacts [3]. This operation will generate a pulse with a duration of 0.25-0.45 cycles which reduces the injected energy during fault testing and prevents damage to equipment. Starting from one phase, in case that the pulse test does not find a fault, having high pulse voltage and low current magnitude, the device initiates the close operation in that phase. The process will continue until either all three phases are closed or a fault (high fault current) is detected in one of the phases. In this case, the operation stops and all poles are open at the end of the pulse testing.

Figs. 1 and 2 show the response to a temporary and a permanent fault with pulse testing technology. Note that SS and LS stand for source-side and load-side measurements, respectively. In case of a permanent fault, both the pulse and the inverse pulse in phase $C$ detect high fault current level and the device will keep all the contacts open. From Fig. 2, the peak fault current of $1400$ A can be observed as the result of pulse testing. This magnitude may vary between 800-1500 A depending on fault position and feeder characteristics.

### B. Time Domain Analysis

The analysis of each individual signal in time domain can be used to provide basic information about the event. In this paper, root-mean-square (rms) values are used to detect changes in the status of each pole (1 = close, 0 = open) and screen the data to identify windows containing possible pulses. The rms of measurement signal $x$ for time window of $m$ can be calculated as follows

$$x_{rms} = \sqrt{\frac{1}{m}\sum_{n=1}^{m} |x_n|^2} \quad (1)$$

The flowchart of the proposed algorithm is shown in Fig. 3. This analysis composed of calculating the rms of voltage and current signals for each phase, finding the correlation between these measurements, and finally comparing the rms values against predefined thresholds. Algorithm 1 first detects the fault-on and pole status flags by comparing the rms (64 samples) value of measurement signals with predefined thresholds $I_{s,th}$ for current and $V_{s,th}$ for voltage measurements.

This algorithm also selects manageable data windows of interest (might contain a pulse) by comparing the rms (16 samples) value of the remaining signals to the pulse thresholds of $I_{p,th}$ and $V_{p,th}$ for current and voltage measurements, respectively. As mentioned in the introduction, pulse detection is necessary to identify and visualize the operation of pulse-reclosers. However, it is sometimes difficult to detect these pulses correctly just based on the screening algorithm. The observation indicates that more candidate pulse windows would be picked up by this algorithm and there is a need to consider the pulse signature to verify the candidates. In the next section, a sparse representation technique is applied as the $2^{nd}$ classification method to provide detailed information about the pulse signature.

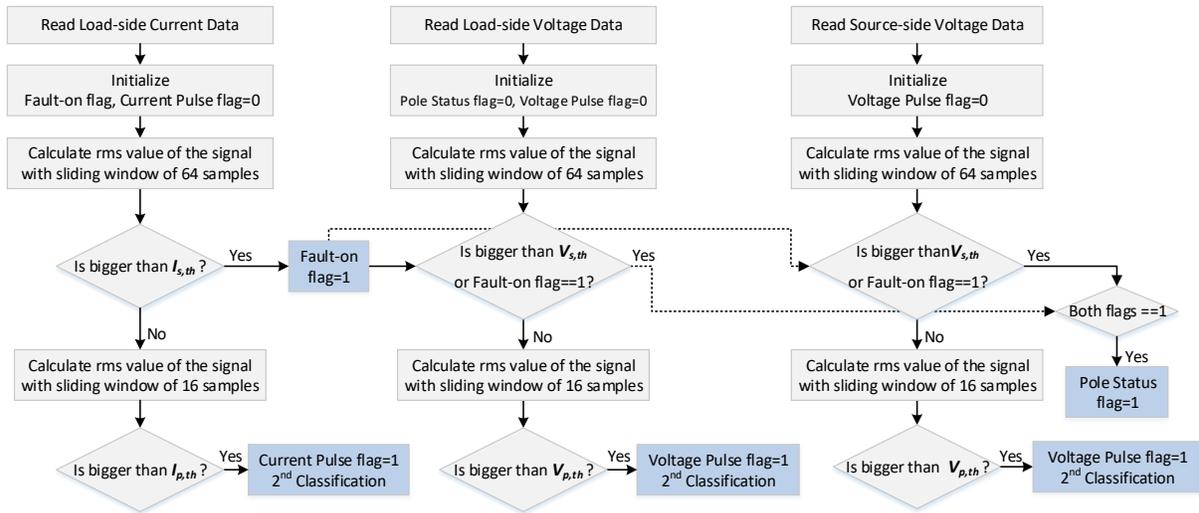

Fig. 3. Flowchart of algorithm 1 for the integrated time domain analysis.

## III. SPARSE REPRESENTATION

Sparse representation is a well-accepted method to describe signals because natural signals are often sparse when the description is done in a proper space of a specific basis. These set of bases that describe the space for signal representation is called a dictionary in the signal processing literature. Each column of the dictionary is called an atom. Modeling data in this scheme is based on the ability to represent input data as linear combinations of a few dictionary elements. In this work, we learn a dictionary to perform best on a training set.

### A. Notation

Assume a set of $N$ inputs denoted by $\{x_i\}_{i=1}^{N}$, where $x_i$ is a vector in $\mathbf{R}^n$ with $n$ being the dimension. We consider the dictionary with $p$ elements in $\mathbf{R}^{n \times p}$ as $D = [d_1, \ldots, d_p]$ where each element or atom $d_k$ is a vector of size $n$. The dictionary $D$ decomposes data $x_i$, to a sparse coefficient vector $\alpha_i \in \mathbf{R}^p$.

### B. Framework

A common approach in statistics, machine learning, and signal processing to learn a vector of parameters $\alpha$ in $\mathbf{R}^p$ requires minimizing a convex function $f : \mathbf{R}^p \to \mathbf{R}_+$. This function measures how well $\alpha$ fits some data. We consider the function $f$ to be smooth, differentiable with a Lipschitz continuous gradient. The criteria to choose the function $f$ depends on the application. Following the concept of empirical risk [10], we measure the quality of the data-fitting with a square loss function: $f \triangleq \|x_i - D\alpha_i\|_2^2$. A regularization term $\Omega : \mathbf{R}^p \to \mathbf{R}_+$, enforces the prior knowledge, here sparsity as $\Omega \triangleq \|\alpha_i\|_{\ell_1}$. Following basis pursuit, given the dictionary $D$, we obtain sparse codes $\alpha_i$ as

$$\underset{\alpha_i \in \mathbf{R}^p}{\mathrm{argmin}} \; \frac{1}{2} \sum_{i=1}^{N} \|x_i - D\alpha_i\|_2^2 + \lambda \|\alpha_i\|_{\ell_1} \quad (2)$$

where the scalar $\lambda \geq 0$ is known as the regularization parameter and it contains the impact of $\Omega$. This optimization can be solved efficiently using interior-point method as described in [11]. In basis pursuit, signal $x$ in $\mathbf{R}^n$ is represented as a linear combination of $p$ columns $d_1, \ldots, d_p$ of the dictionary $D \in \mathbf{R}^{n \times p}$. The dictionary $D$ is either fixed [12] or made from learned representations as in [13]. In (2), without any regularization over $D$, the $\alpha_i$ coefficients would be small. That is, we enforce the set $\mathcal{D}$ to be the set of matrices whose columns are bounded by a unit 2-norm ball as follow

$$\mathcal{D} \triangleq \{D \in \mathbf{R}^{n \times p} \; \text{s.t.} \; \forall k \in \{1, 2, \ldots, p\}, \|d_k\|_2^2 \leq 1\} \quad (3)$$

### C. Signature Dictionary Construction

We largely follow the sparse representation classification (SRC) introduced for application of face recognition in [14]. Here, the task is to discriminate the target (pulse) from the background; hence, it is a binary classification problem where $c \in \{-1, +1\}$ and $c = +1$ refers to the target class while $c = -1$ refers to the background. The goal of the training phase is to obtain the dictionary $D$. There are $N_{+1}$ signals from the target and $N_{-1}$ signals from the background which results in $N_{-1} + N_{+1}$ total training data. We manually select $N_{+1}$ samples with time window $w$ with a pulse inside, which we refer to as $X_{+1} \in \mathbf{R}^{w \times N_{+1}}$. Similarly, we take $N_{-1}$ samples with the same time window $w$ that do not have the target inside, referred to as $X_{-1}$ with the size $\mathbf{R}^{w \times N_{-1}}$.

Typically pulses and inverse-pulses might have different characteristics. Meanwhile, pulses in voltage or current measurements also have different signatures as shown in Fig. 4. These pulses have been separated in each measurement, normalized to have a unit 2-norm and then used to build the dictionary. The dictionary of the target, $D_{+1}$ with 10 atoms, is obtained by running two separate k-means with $k = 5$ clusters on the target pulse and inverse-pulse signatures with a time window of 180 samples (almost 3 cycles). The dictionary construction for load-side voltage is shown in Fig. 5. We did the same to obtain the dictionary of the background, $D_{-1}$ with 40 atoms. The final dictionary with $p = 50$ atoms is made by combining all class specific dictionaries, $D = [D_{+1}, D_{-1}]$. This strategy obtains a dictionary with known labels.

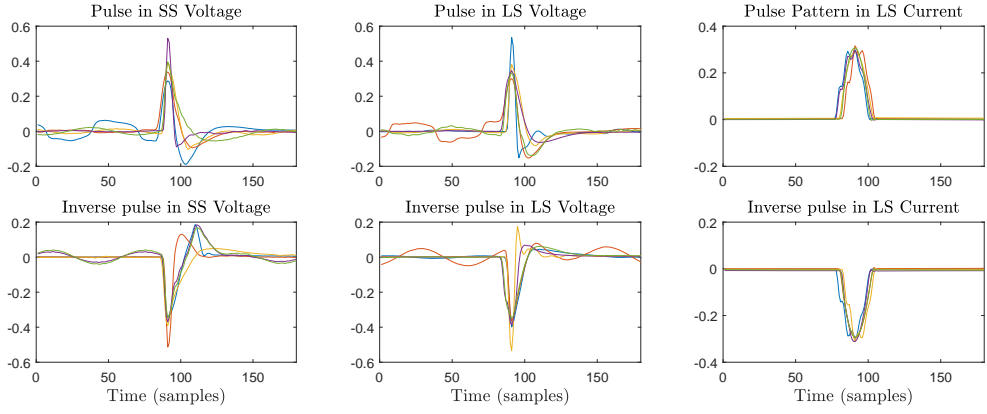

Fig. 4. Normalized pulse and inverse-pulse signature used to build the target dictionary $D_{+1}$ for different measurements.

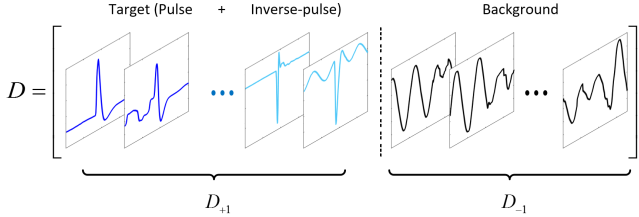

Fig. 5. Illustration of the dictionary $D$ construction for load-side voltage.

### D. Testing Step

The output of training step is the dictionary $D = [D_{+1}, D_{-1}]$. Given a signal $x_q$ with size $w$, the task is to identify if it is a target. We assume that the test signal from target class can be represented using atoms that belong to the target (the $D_{+1}$ part of the dictionary $D$). In other words, the test sample $x_q$ belongs to the target class, if it can be approximated using atoms of the target dictionary as $x_q \approx D\alpha_q$, where $\alpha_q$ is the sparse vector of the signal $x_q$ obtained from (2). For $x_q$ to be the target class, its $\alpha_q$ should be zero everywhere except for atom indices that belong to the target class, i.e. $\alpha_q = [\alpha_{+1}^\top, 0^\top]^\top$. That is, to reconstruct the query using the minimum number of atoms, which is equal to search for sparse vector $\alpha_q$ to reconstruct the test signal $x_q$ by minimizing the $\ell_1$-norm as in (2).

Assuming the query to belong to the target class, the vector $\delta_c(\alpha_q)$ in $\mathbf{R}^p$ is zero everywhere except entries that are associated with the target class: $\delta_c(\alpha_q) = [\alpha_{+1}^\top, 0^\top]^\top$. The test data is approximated as: $\hat{x}_q = D\delta_c(\alpha_q)$. In the next step, the data may be assigned to the class label that can reconstruct the query with least reconstruction error

$$\hat{c} = \underset{c}{\operatorname{argmin}} \ \|x_q - D\delta_c(\alpha_q)\|_2 \quad (4)$$

In this work, the data will be assigned to a target class if the reconstruction error is lower than the background class and $c^* = 1 - \hat{c}$ is bigger than a predefined threshold.

## IV. EXPERIMENTS

To further illustrate the use of this methodology, this section evaluates the performance of the proposed method on data collected from 12kV distribution network of the Electric Power Board (EPB), a distribution company in Chattanooga TN. The data is taken from 20 sites randomly, and is divided equally to test and training sets. Based on their settings, the pulse-reclosers record three-phase source/load-side voltages and load-side currents with 64 samples per cycle. After an event is recorded, the data was transferred to the local server for storage and further analysis.

A program in MATLAB[2] is developed to analyze the event data and visualize the operation of these devices. The parameters in step 1 are chosen as $V_{s,th} = 4.0$ kV, $V_{p,th} = 2.8$ kV, $I_{s,th} = 1.0$ kA and $I_{p,th} = 0.9$ kA. These parameters are chosen throughout extensive studies based on the minimum rms values of voltages, fault currents and pulse events. The regularization parameter is selected as $\lambda = 0.2$ and the pulse reconstruction threshold is $0.4$ to ensure the correct pulse detection. Note that regularization parameter $\lambda$ adjust the trade-off between sparsity and reconstruction error. This implies that choosing a large value for $\lambda$ may lead to increasing the reconstruction error and if $\lambda$ is too small, the representation will not be sparse anymore. In this section, three case study scenarios representing pulse in different measurements are analyzed and a summary of performance is provided.

### A. Case 1: Pulse testing in upstream device

Case 1 is a pulse testing event in the upstream device as shown in Fig. 6. In the sequence of events, the device which records this data and the one upstream both trip open. The upstream device performs pulse testing and several pulse candidates are detected (marked with black boxes) in both load-side and source-side voltages. Candidate pulse windows are then being further analyzed using the sparse representation technique to verify the pulse candidates.

An example of the sparse coefficient vector and reconstruction error are shown in Fig. 7. From the top sub-figure, it is clear that the candidate window does not contain a pulse waveform and has a higher $c^*$ (lower reconstruction error) using background signals. The bottom sub-figure represents the case where the window contains a pulse. The analysis results are also shown in the last sub-figure in Fig. 6. It can be seen that all the poles are tripped open (pole status = 0) before $t = 5$ and there exist a pulse in phase $C$ and $B$ before $t = 15$ and $t = 25$ cycle. This simple visualization presents useful information and indicates that the proposed method can successfully detect the pulse and status of each pole.

---

[2]MATLAB is a registered trademark of The Mathworks, Inc.

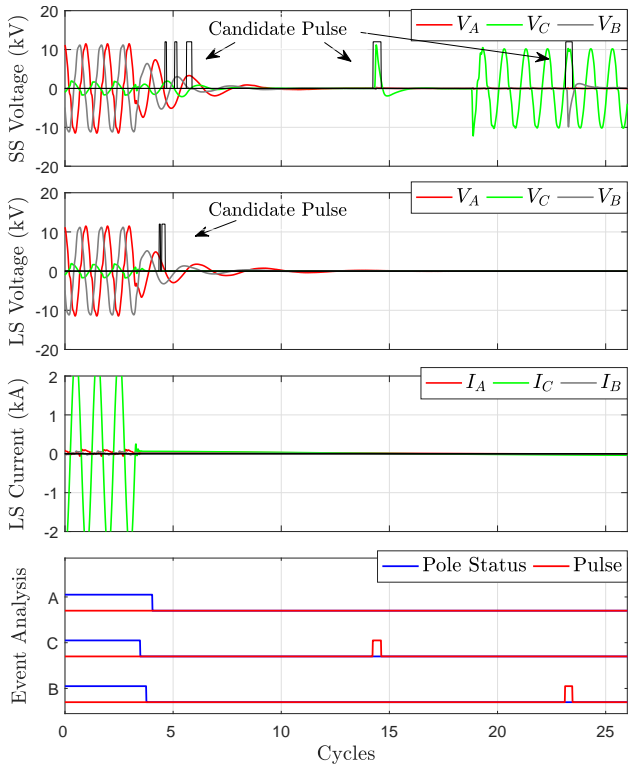

Fig. 6. Event analysis result for case 1 where upstream device is performing pulse testing.

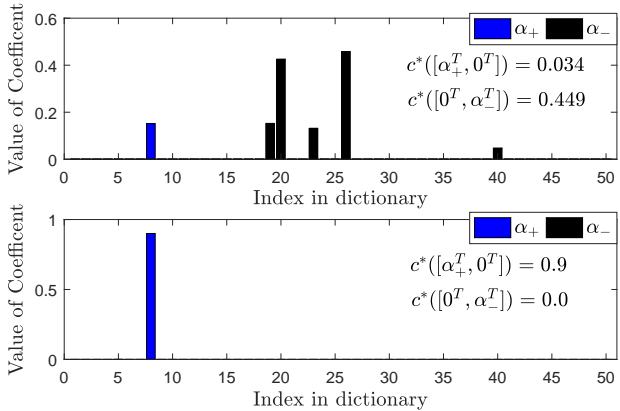

Fig. 7. Sparse coefficient $\alpha$ in case candidate window does not contain a pulse (top) and the case where candidate window contains a pulse (bottom).

### B. Case 2: Pulse testing into a temporary fault

In case 2-1, the pulse-recloser initially trips open and the source-side voltage is available as shown in Fig. 8. The device performs pulse testing and closes pole $B$. This causes the pulse-recloser to see voltage on the other two phases which then offsets the pulse. One possible reason is having a delta transformer downstream. This offset alters the signature of pulse in phase $A$ and $C$. However, our method still correctly classifies these pulses from background signals. The last sub-figure visualizes the event analysis results and a successful pulse testing operation can be seen.

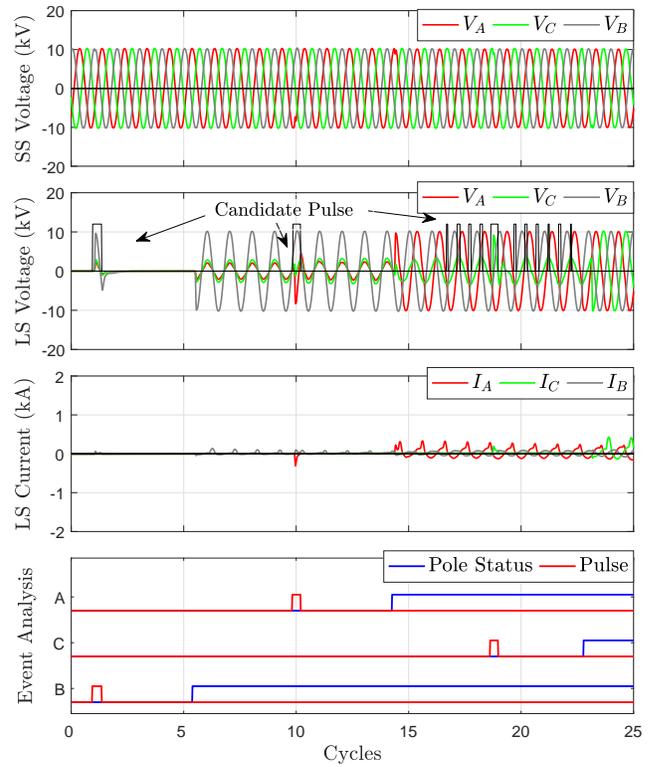

Fig. 8. Event analysis result for case $2-1$ where the device is performing pulse testing.

Case 2-2 is another pulse testing case into a temporary fault as shown in Fig. 9. In this case, high inrush currents misleading algorithm 1 to indicate candidate pulse window. Finally, the event analysis results are shown in the last sub-figure in Fig. 9. It can be seen that further analysis using sparse representation techniques classifies all the candidates in the LS voltage as correct pulse and classifies the inrush currents as background signals, which do not contain any pulse signatures.

### C. Case 3: Pulse testing into a permanent fault

In case 3, the pulse-recloser is performing pulse testing into a permanent fault as shown in Fig. 10. Three candidate pulse windows are detected, one on the load-side voltage and two in the load-side current measurements. The sparse representation verifies the existence of a pulse in all candidate windows. From the event analysis visualization, it can be concluded that a line to ground (L-G) permanent fault in phase $C$ leads to a failed pulse testing operation. As a result, the operation stops and all poles are open at the end of the pulse testing.

### D. Detection and classification performance

Recorded data from other pulse-reclosers containing 120 events and 143 pulses are analyzed using the proposed method. Table I provides a summary of the performance of the proposed method for pulse detection and classification. We use two metrics to evaluate the performance. Correct pulse detection rate is the number of correctly identified pulses (detected by the proposed method) expressed as a percentage of the real number of pulses. Similarly, false pulse detection rate represents the ratio between the number of detected pulses that are incorrect and the true number of pulses. It can be concluded that the proposed method has a very high accuracy. Furthermore, it can be seen that there are more pulses in the LS voltage in compare to the current measurements which confirms that most events caused by temporary faults.

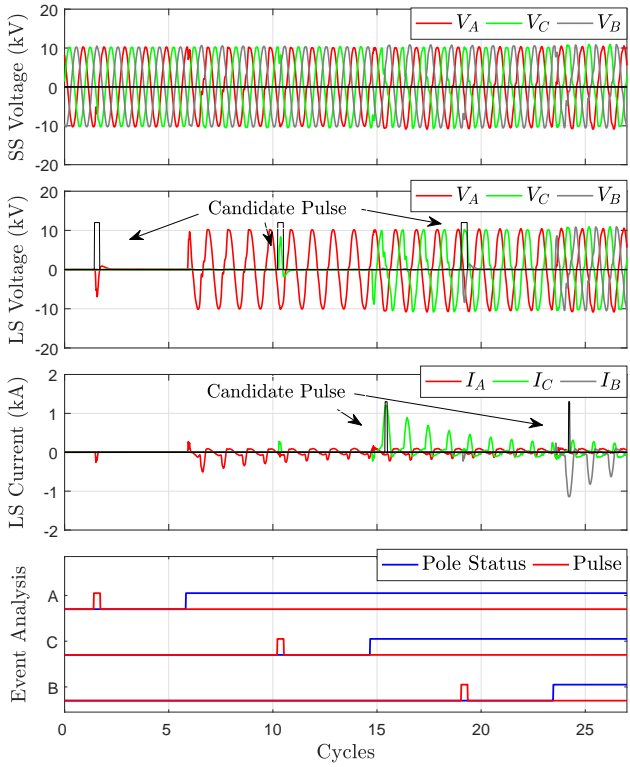

Fig. 9. Event analysis result for case $2-2$ where the device is performing pulse testing in presence of high inrush currents.

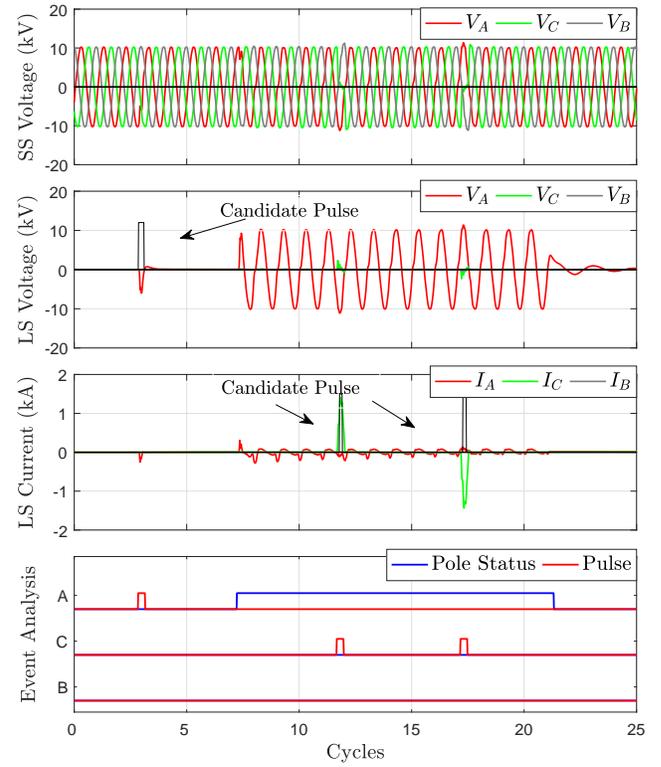

Fig. 10. Event analysis result for case 3 where the device is performing pulse testing into a line to ground permanent fault.

TABLE I
QUANTITATIVE EVALUATION USING REAL RECORDED DATA

|  | No. of Pulses | Correct Pulse Detection (%) | False Pulse Detection (%) |
| --- | --- | --- | --- |
| SS Voltage | 13 | 84.61 | 0.00 |
| LS Voltage | 108 | 97.22 | 0.01 |
| LS Current | 22 | 90.90 | 0.00 |

## V. CONCLUSION

This paper proposes a methodology based on time domain analysis and sparse representation technique for event analysis of pulse-reclosers. An algorithm is developed to screen the data to identify the status of each pole and tag time windows with a possible pulse event. The selected time windows are further analyzed and classified using a sparse representation technique through solving an $\ell_1$-regularized least-square problem. From the case studies presented, it can be observed that the initial step is successful in terms of finding the status of each pole. However, this step detects more pulse candidate windows, which must then be classified using sparse representation. Field recorded data from a distribution system is used to verify the proposed approach and a simple visualization is proposed to present useful information about the events.